\documentclass[a4paper]{jpconf}
\usepackage{graphicx}

\def\lsim{\raise0.3ex\hbox{$\;<$\kern-0.75em\raise-1.1ex\hbox{$\sim\;$}}}
\def\gsim{\raise0.3ex\hbox{$\;>$\kern-0.75em\raise-1.1ex\hbox{$\sim\;$}}}
\def\barr{\begin{eqnarray}}
\def\earr{\end{eqnarray}}
\def\beq{\begin{equation}}
\def\eeq{\end{equation}}

\newcommand{\ba}{\begin{array}{c}}
\newcommand{\ea}{\end{array}}

\def\nue{{\nu_e}}
\def\nuebar{{\bar{\nu}_e}}

\def\nux{{\nu_x}}

\def\pbar{\bar{p}}

\begin{document}
\title{Physics potential of future supernova neutrino observations}

\author{Amol Dighe}

\address{Tata Institute of Fundamental Research, Homi Bhabha Road,
Colaba, Mumbai 400005, INDIA}

\ead{amol@theory.tifr.res.in}

\begin{abstract}
We point out possible features of neutrino spectra from a future galactic
core collapse supernova that will enhance our understanding of neutrino 
mixing as well as supernova astrophysics.
We describe the neutrino flavor conversions inside the star, emphasizing
the role of ``collective effects'' that has been appreciated and
understood only very recently.
These collective effects change the traditional predictions of flavor 
conversion substantially, and enable the identification of neutrino
mixing scenarios through signatures like Earth matter effects. 
\end{abstract}

\section{Introduction}

Time dependent energy spectra of $\nue$ and $\nuebar$ from a future
galactic supernova (SN) can be decoded to obtain information on
primary neutrino spectra, the neutrino mixing scheme as well as 
densities encountered by the neutrinos along their path. 
In particular, these spectra enable us to probe extremely small
$\theta_{13}$ values and distinguish between normal and inverted 
mass hierarchies \cite{dighe-smirnov}.
On the other hand, the neutrino signal provides important
astrophysical information that allows us to point at the SN in
advance and even track the shock wave while it is still inside
the mantle (see \cite{nufact07} for a recent overview).

The only SN observed in neutrinos till now, SN1987A, yielded only 
$\sim$20 events. Though it confirmed our understanding of the SN 
cooling mechanism \cite{book}, the number of events was too small to say 
anything concrete about neutrino mixing (see \cite{ls-sn87} and 
references therein).
On the other hand, if a SN explodes in our galaxy at $\sim 10$ kpc from 
the Earth, we expect ${\cal O}(10^4)$ events at Super-Kamiokande (SK) 
through the inverse beta decay process $\nuebar + p \to n + e^+$
\cite{raj}.
This process, dominant at any water Cherenkov or scintillation 
detector, will be instrumental in determining the $\nuebar$ spectrum.
The number of events will be higher by an order of magnitude 
at the planned megaton scale water Cherenkov detectors 
\cite{Autiero:2007zj,Jung:1999jq,Nakamura:2003hk,Suzuki:2001rb},
while large scintillation detectors \cite{lena} 
will determine the neutrino energy with much better precision.
Even a gigaton ice Cherenkov detector like IceCube \cite{halzen,halzen2}, 
though incapable of detecting SN neutrinos individually, can determine
the total $\nuebar$ luminosity as a function of time.  
In order to measure the $\nue$ spectrum cleanly, one needs a large 
liquid Ar detector, with the relevant process
$\nu_e + ~^{40}{\rm Ar} \to ~^{40}{\rm K}^* + e^-$,
with ${\cal O}(10^4)$ events expected at a 100 kt detector 
\cite{GilBotella:2003sz,GilBotella:2004bv}.
With such a significant statistics, one would be able to
reconstruct the $\nu_e$ and $\nuebar$ spectra, and extract
the information encoded therein.

The traditional analysis of neutrino flavor conversions \cite{nu04}
neglected the neutrino-neutrino interactions, which are now known to be
significant near the neutrinosphere owing to the large neutrino
density. The last couple of years have seen significant progress
in our understanding of these neutrino refraction effects, also
known as the collective effects. In this talk, we shall concentrate
on these recent exciting developments, clarify our current
understanding of SN neutrino flavor conversions, and obtain the
characteristics of the observed neutrino fluxes.
Finally, we shall point out the distinctive features to look for
in the neutrino spectra that may be able to settle some of the
unresolved problems in neutrino physics and astrophysics.
We restrict ourselves to standard three-neutrino mixing.

\section{Flavor conversions of supernova neutrinos}

\subsection{Neutrino production and primary spectra}

Before the core collapse, neutrinos of all species are trapped
within the star inside their respective ``neutrinospheres''
around $\rho \sim 10^{10}$g/cc.
When the iron core reaches a mass close to its Chandrasekhar limit,
it becomes gravitationally unstable and collapses.
A hydrodynamic shock is formed when the matter reaches nuclear
density and becomes incompressible.
When the shock wave passes through the
$\nue$ neutrinosphere, a short $\nue$
``neutronization'' burst is emitted, which
lasts for $\sim$10 ms.
The object below the shock wave, the ``protoneutron star,'' then
cools down with the emission of neutrinos of all species,
over a time period of $t \sim 10$ s \cite{book}.
The eventual explosion of the star involves the stalling of the original
shock wave, its revival by the trapped neutrinos, and
a ``delayed'' explosion where large scale convections play
an important role \cite{bethe,janka}.

The SN core acts essentially like a neutrino black-body source
with flavor dependent fluxes. Since the fluxes are almost 
identical for $\nu_\mu$ and $\nu_\tau$ ($\bar{\nu}_\mu$ and $\bar{\nu}_\tau$), 
it is convenient to denote these species collectively 
by $\nu_x$ ($\bar{\nu}_x$).
The ``primary fluxes'' $F^0_{\nu_\alpha}$ are parametrized by the 
total number fluxes $\Phi_0(\nu_\alpha)$, average energies 
$\langle E_0(\nu_\alpha) \rangle$, and the ``pinching
parameters'' that characterize their spectral shapes ~\cite{keil1}.
The values of the parameters are highly model dependent,
as can be seen from table~\ref{tab:models}.

\begin{table}[h]
\begin{center}
\caption{\label{tab:models}
Typical neutrino Flux predictions in the 
Livermore \cite{livermore} and Garching
\cite{garching} models. Average energies are given in MeV.}
\begin{tabular}{lcccccc}
\br
{Model} & 
${\langle E_0(\nue) \rangle}$ &
${\langle E_0(\nuebar) \rangle}$ &
${\langle E_0(\nux) \rangle}$ &
${\langle E_0(\bar{\nu}_x) \rangle}$ &
$\frac{\Phi_0(\nue)}{\Phi_0(\nux)}$ &
$\frac{\Phi_0(\nuebar)}{\Phi_0(\bar{\nu}_x)}$\\
\mr
{Livermore} & 12 & 15 & 24 &
{24 }& {2.0}&{1.6} \\
{Garching} & 12 & 15 & 18 &
{18} & {0.8 }& {0.8} \\
\br
\end{tabular}
\end{center}
\end{table}

In the light of the large model dependence in the primary neutrino
spectra, it is imperative to look for signals at these detectors
that are independent of the details of the primary spectra, but
depend on distinctive signatures of neutrino mixing schemes.
In order to do this, one needs to analyze the neutrino flavor conversions 
during their propagation outwards from the neutrinospheres.

\subsection{Collective effects at large neutrino densities}
\label{collective}

The neutrino and antineutrino densities near the neutrinosphere 
are extremely high ($10^{30-35}$ per cm$^3$), which
make the $\nu-\nu$ interactions in this region significant
\cite{duan-fuller-carlson-qian-0606616,duan-fuller-carlson-qian-0608050}.
Such a dense gas of neutrinos and antineutrinos is coupled to
itself, making its time evolution nonlinear
\cite{thompson-mckellar-PLB259,raffelt-sigl-NPB406,thompson-mckellar-PRD49}.
The flavor changing terms are sizeable, as a result
significant flavor conversions can occur.
The distinctive features in the flavor evolution of such 
a relativistic gas have been identified in
\cite{samuel-PRD48,kostecky-samuel-9506262,pantaleone-PRD58,samuel-9604341}.

Analytic studies of collective effects reveal a rich phenomenology
of flavor conversions, with phenomena that are completely
distinct from the traditional vacuum or MSW oscillations.
These include ``synchronized oscillations''
\cite{pastor-raffelt-semikoz-0109035}, where
$\nu$ and $\bar{\nu}$ of all energies oscillate with the same frequency,
``bipolar oscillations'' 
\cite{hannestad-raffelt-sigl-wong-0608095,duan-fuller-carlson-qian-0703776}
that correspond to pairwise conversions
$\nu_e \bar{\nu}_e \leftrightarrow \nu_y \bar{\nu}_y$
(where $\nu_y$ is a specific linear combination 
of $\nu_\mu$ and $\nu_\tau$),
and ``spectral split'' 
\cite{raffelt-smirnov-0705.1830,raffelt-smirnov-0709.4641},
where $\bar{\nu}_e$ and $\bar{\nu}_x$ spectra
interchange completely (barring some possibly non-adiabatic 
effects at very low energies \cite{nubar-split}),
whereas $\nu_e$ and $\nu_x$ spectra 
interchange only above a certain critical energy $E_c$.

The dynamics in three generations can be factorized into 
a superposition of multiple two-flavor phenomena with hierarchical 
frequencies, and can be visualized in terms of the so-called
``${\bf e}_3$--${\bf e}_8$'' triangle diagrams \cite{threeflavor}.
In addition, some new three flavor effects also emerge:
for example in early accretion phase, large $\mu$-$\tau$ matter potential 
causes interference between MSW and collective effects, 
which is sensitive to deviation of $\theta_{23}$ from maximality
\cite{mutau-refraction}.

The dependence of the flavor evolution on the direction of
propagation of the neutrino may give rise to direction-dependent
evolution 
\cite{duan-fuller-carlson-qian-0606616,duan-fuller-carlson-qian-0608050},
or to decoherence effects
\cite{pantaleone-PRD58,raffelt-sigl-0701182,EstebanPretel:2008ni}.
Such multi-angle effects, extensively studied numerically
in \cite{fogli-lisi-marrone-mirizzi-0707.1998,Duan:2008eb}, 
have recently been interpreted in terms of 
``neutrino flavor spin waves'' \cite{Duan:2008fd}.
However, for a realistic asymmetry between neutrino and antineutrino
fluxes, such multi-angle effects are likely to be small
\cite{fogli-lisi-marrone-mirizzi-0707.1998,estebanpretel-pastor-tomas-raffelt-sigl-0706.2498}
and a so-called ``single-angle'' approximation  can be used.
Even for non-spherical geometries, one can study the evolution
along stream lines of neutrino flux, as long as 
coherence is maintained \cite{nonspherical}.

Though the inherent nonlinearity and the presence of multi-angle
effects make the analysis rather complicated, the outcome for
the flavor conversions turns out to be rather straightforward,
at least when the number fluxes of $\nue,\nuebar$ and $\nux$ 
follow an hierarchy $N_\nue > N_\nuebar > N_\nux$.
The propagation of the neutrinos can be rather cleanly separated 
into regions where various collective effects dominate individually
\cite{threeflavor,fogli-lisi-marrone-mirizzi-0707.1998},
and hence the neutrinos experience these effects sequentially.
Just outside the neutrinosphere synchronized oscillations occur,
which however cause no significant flavor conversions since
the mixing angle is highly suppressed owing to large matter density.
If the hiearchy is inverted, bipolar oscillations follow,
which prepare the neutrinos for the eventual spectral split.
Thus when the neutrinos emerge from the region where collective effects
dominate, inverted hierarchy simply predicts a complete swap of 
$\bar{\nu}_e$ and $\bar{\nu}_x$ spectra and a swap of 
 $\nu_e$ and $\nu_x$ spectra above a critical energy $E_c$.
In the normal hierarchy, collective effects do not cause
any intermixing of neutrino spectra \cite{threeflavor}.

\subsection{MSW resonances inside the SN and propagation through vacuum}

In iron core supernovae, the collective effects have already become
insignificant when neutrinos enter the MSW resonance regions.
Therefore, traditional flavor conversion analysis can be
applied to the fluxes emerging from the high density region
\cite{nufact07}.
SN neutrinos must pass through two resonance layers: 
the H-resonance layer at
$\rho_{\rm H}\sim 10^3$ g/cc characterized by
$(\Delta m^2_{\rm atm}, \theta_{13})$,
and the L-resonance layer at
$\rho_{\rm L}\sim 10$ g/cc characterized by 
$(\Delta m^2_{\odot}, \theta_{12})$.
The outcoming incoherent mixture of vacuum mass eigenstates 
travels to the Earth without any further conversions,
and is observed at a detector as a combination of primary fluxes
of the three neutrino flavors:
\begin{equation}
F_{\nue}  =  p F_{\nue}^0 + (1-p) F_{\nux}^0  \; , \quad \quad 
F_{\nuebar}  =   \pbar F_{\nuebar}^0 + (1-\pbar) 
F_{\nu_x}^0 ~,
\label{feDbar}
\end{equation}
where $p$ and $\pbar$ are the
survival probabilities of $\nue$  and $\nuebar$ respectively.

The neutrino survival probabilities are governed by
the adiabaticities of the resonances traversed, 
which are directly
connected to the neutrino mixing scheme.
In particular, whereas the L-resonance is always adiabatic and
appears only in the neutrino channel, the adiabaticity of
the H-resonance depends on the value of $\theta_{13}$, and
the resonance shows up in the neutrino (antineutrino) channel
for a normal (inverted) mass hierarchy.
Table~\ref{tab:pbar} shows the survival probabilities
in various mixing scenarios.
For intermediate values of $\theta_{13}$, i.e.
$10^{-5}\lsim\sin^2 \theta_{13} \lsim 10^{-3}$,
the survival probabilities depend on energy as well
as the details of SN density profile
\cite{dighe-smirnov}.
 
\begin{table}[ht]
\begin{center}
\caption{\label{tab:pbar} Survival probabilities for neutrinos, $p$, and
antineutrinos, $\pbar$, in various mixing scenarios, in the cooling phase
when $N_\nue > N_\nuebar > N_\nux$.
The presence or absence of shock wave effects and Earth effects
is denoted by $\surd$ and X respectively. 
The notation $p_1 \parallel p_2$ indicates that the survival probability 
is $p_1$ ($p_2$) for energies below (above) the critical energy $E_c$.
}
\begin{tabular}{llccccccc}
\br
{} &
{  Hierarchy} &  {$\sin^2 \theta_{13}$}  & 
\multicolumn{2}{c}{Survival probability} &
\multicolumn{2}{c}{Shock effects} &
\multicolumn{2}{c}{Earth effects} \\
 & & & 
{  $p$} & {  $\bar{p}$} & {  $\nu_e$} & {  $\bar{\nu}_e$} &
{  $\nu_e$} & {  $\bar{\nu}_e$} \\
\mr
{  A} & Normal & { $\gsim 10^{-3}$}  & 0  & $\cos^2\theta_\odot$ 
& $\surd$  & $\surd$ & X & $\surd$ \\
{  B} & Inverted &  { $\gsim 10^{-3}$} &  
$\sin^2\theta_\odot \parallel 0$ &  $\cos^2 \theta_\odot$ &
X &  $\surd$ & X & $\surd$ \\
{  C} & Normal & { $\lsim 10^{-5}$}  & $\sin^2\theta_\odot$
&  $\cos^2\theta_\odot$ & X
& X & $\surd$ & $\surd$ \\
{  D} & Inverted & {$\lsim 10^{-5}$}  & 
$\sin^2\theta_\odot \parallel 0 $
&  0 & 
X &  X   & X & X \\
\br
\end{tabular}
\end{center}
\end{table}
The scenarios A, B, C and D in table~\ref{tab:pbar}
are the ones that can in principle be
distinguished through the observation of a SN neutrino burst.
Note that all these scenarios are characterized by distinct 
combinations of $p$ and $\pbar$, which was not the case 
when the $\nu$--$\nu$ interactions were neglected, making 
scenarios C and D degenerate \cite{nufact07}.
This is a consequence of the sensitivity of collective effects
to the mass hierarchy even when $\theta_{13}$ is
as small as $10^{-10}$ \cite{fuller-split,earth-hierarchy}.

\subsection{Propagation through the shock wave near MSW resonances}
  
The passage of the shock wave through the H-resonance 
($\rho \sim 10^3$ g/cc) a few seconds after the core bounce may
break adiabaticity, thereby modifying the flavor evolution
of neutrinos that are emitted during this time interval
~\cite{fuller-shock,takahashi,ls1,lisi-shock,revshock,phase}.
Such a situation is possible, even in principle,
only in the scenarios indicated in table~\ref{tab:pbar}.
As a result, the identification of shock wave effects provides an
important input for distinguishing between neutrino mixing schemes.

The shock wave effects can be diluted by stochastic density
fluctuations \cite{stochastic} or turbulence \cite{friedland}.
However, a recent hydrodynamic simulation \cite{brockman}
suggests that some of the shock wave effects survive
in spite of these smearing factors.

\subsection{Oscillations inside the Earth matter}
\label{earth}

If the neutrinos travel through the Earth before reaching the detector, 
they undergo flavor oscillations and the survival probabilities change
\cite{cairo,ls2,sato}.
This change however occurs only in those scenarios indicated
in table~\ref{tab:pbar}. The presence or absence of Earth
effects in the neutrino or antineutrino channels therefore would
help in identifying some of the mixing schemes.

\section{Smoking gun signals in neutrino spectra}

\subsection{Direction of neutrino arrival}
\label{pointing}

Since neutrinos are expected to arrive hours before the optical
signal from the SN, the neutrino burst serves as an early warning
\cite{snews} to the astronomy community.
Being able to determine the position of the SN in the sky is
also crucial for determining the Earth crossing
path for the neutrinos in the absence of the SN observation
in the electromagnetic spectrum.

A SN may be located through the directionality of
the $\nu e^- \to \nu e^-$ elastic scattering events
in a water Cherenkov detector~\cite{beacom,ando}.
The large but nearly isotropic $\nuebar p \to n e^+$ background
can be removed by the addition of a small amount of gadolinium
to water so that neutrons from the inverse beta decay are 
tagged \cite{gadzooks,pointing}.

\subsection{Suppression of $\nue$ in the neutronization burst}

Since the primary signal during the neutronization burst is 
pure $\nu_e$, and the model predictions for the energy and luminosity
of the burst are fairly robust \cite{ricard-neutronization},
the observation of the burst signal gives direct information
about the survival probability of $\nu_e$. This probability is
${\cal O}(\theta_{13}^2)$ in scenario A and $\sin^2 \theta_\odot$
in all the other scenarios \cite{dighe-smirnov}. 
(Note that collective effects do not affect this scenario even for 
inverted hierarchy, since the absence of $\nuebar$ implies that 
bipolar oscillations do not develop.)
Thus, the vanishing of $\nu_e$ burst
would be a smoking gun signal for scenario A.
In order to be able to separate the $\nu_e$ burst from the accretion 
phase signal, time resolution of the detector is crucial
\cite{ricard-neutronization}.

An O-Ne-Mg supernova offers another interesting possibility.
For such a light star, the MSW resonances may lie deep inside
the collective regions during the neutronization burst, when
the neutrino luminosity is even higher.
In such a situation, neutrinos of all energies undergo the
MSW resonances together, with the same adiabaticity. As long
as this adiabaticity is nontrivial, one gets the ``MSW-prepared
spectral splits'', two for normal hierarchy and one for inverted
hierarchy \cite{fuller-split1,fuller-split2,onemg-doublesplit}.
The positions of the critical energies for these splits can be
predicted from the primary spectra \cite{onemg-doublesplit}.
The splits imply $\nu_e$ suppression that is stepwise in energy.
Such a signature may even be used to identify the O-Ne-Mg
supernova, in addition to identifying the hierarchy.

\subsection{Shock wave effects}
\label{shock}

Shock wave effects result in sharp changes in 
characteristics of the observed spectra that occur 
for a very short time ($\sim 1$ s) while the shock wave is
passing the H resonance 
\cite{fuller-shock,takahashi,ls1,lisi-shock,revshock,phase}.
Robust observables like the number of events, average energy,
or the width of the spectrum may display dips or peaks 
for short time intervals due to these effects.
If a reverse shock is also present, the above features become 
double-dips or  double-peaks \cite{revshock},
which are difficult to be mimicked by uncertainties in
the time evolutions of neutrino fluxes.
The positions of the dips or peaks in the number of events at different 
neutrino energies would also allow one to trace the shock
propagation while it is in the mantle around densities of
$\rho \sim 10^3$ g/cc \cite{revshock}.

For an iron core SN, a positive identification of any of the 
above shock effects 
in $\nu_e$ spectrum shortlists scenarios A and B, whereas
shock effects in $\nuebar$ identify scenario A.

If light sterile neutrinos exist, they 
may leave their imprints in the shock wave
\cite{choubey-ross1,choubey-ross2}.
For an O-Ne-Mg supernova, passage of the shock wave through
the sharp density profile at the resonance leads to
distinctive effects \cite{lunardini-onemg}.

\subsection{Spectral split in $\nu_e$}
                                              
Table~\ref{tab:pbar} shows that the spectral split in 
neutrinos is absent (present) in the normal (inverted) hierarchy.
Such a feature would in principle be visible in the $\nu_e$
spectrum as a sharp jump at the critical energy $E_c$
\cite{fuller-split}. 
This would happen even for values of $\theta_{13}$
as low as $10^{-10}$, since the split is adiabatic even at such
low values \cite{raffelt-smirnov-0709.4641,earth-hierarchy}. 
The spectral split,
which can in principle be observed at a liquid Ar detector,
 is therefore a smoking gun signal for inverted hierarchy.

However, the sharp split in $p$ is smeared to some extent by the
multi-angle effects 
\cite{fogli-lisi-marrone-mirizzi-0707.1998}.
Moreover, for all the
reasonable values of model parameters for primary spectra,
$E_c$ is less than 10 MeV. At such energies, the low cross section,
finite resolution of the detector, 
and the small difference in the $\nu_e$ and $\nu_x$ spectra 
make the observation of the split a challenging task \cite{liqAr}.

\subsection{Earth matter effects}

Earth matter effects can be identified by the comparison of signals 
at two detectors, only one of which is shadowed by the earth.
This could be achieved through the $\nu_e$ spectra 
at two large water Cherenkov detectors \cite{earth-hierarchy}
or through the time dependent ratio of luminosities
at IceCube and Hyper-Kamiokande \cite{ice-hyper}. 
As can be seen from table~\ref{tab:pbar}, when
$\theta_{13}$ is small, Earth effects on antineutrinos
are present (absent) for normal (inverted) hierarchy 
\cite{earth-hierarchy}.

The Earth effects can be identified even at a single detector
as long as it is capable of determining the neutrino energy.
The measurement of the Fourier power spectrum of the
``inverse energy'' spectrum of $\nuebar$ \cite{fourier}
gives peaks (multiple ones if the neutrinos traverse the Earth core)
whose positions are independent of the primary neutrino 
spectra, that reveal the presence of earth matter effects
\cite{corewiggles}.
Energy resolution of the detector plays a crucial role in 
observing this signal, and a smaller scintillation detector
may compete with a large water Cherenkov \cite{corewiggles}
for detecting Earth effects on the $\nuebar$ spectrum.

Earth effects on $\nue$ spectrum may be identified at 
a liquid Ar detector \cite{liqAr}, which would identify scenario C
positively, as can be seen from table~\ref{tab:pbar}.

\section{Concluding remarks}

Supernova neutrinos can probe extremely small values of $\theta_{13}$
and can determine the neutrino mass hierarchy at $\theta_{13}$ as
low as $10^{-10}$, thanks to
collective effects and MSW resonances inside the star.
Smoking gun signals of neutrino mixing scenarios can be 
independently obtained through observations like
the suppression of neutronization burst,
time variation of the signal during shock wave propagation,
and Earth matter effects.
SN neutrinos also enable pointing at the SN
in advance of the optical signal, and tracking the shock 
wave while it is still inside the mantle.

A future galactic SN is therefore expected to yield a rich
harvest of scientific information for neutrino oscillation physics and
SN astrophysics. Though this is a rare phenomenon, occurring only a 
few times per century in a typical galaxy,
we must make the best of it by being ready with suitable detectors.

\ack

I would like to thank Basudeb Dasgupta, Alessandro Mirizzi and
Georg Raffelt for fruitful collaborations, the organisers of
Neutrino 2008 for their hospitality, and the Max Planck - India
Partnergroup for financial support.

\section*{References}

\end{document}